\documentclass[prl,twocolumn,showpacs,floatfix]{revtex4}

\usepackage{graphicx}
\usepackage{amsmath}

\begin{document}

\title{Three and Four-Body Interactions in Spin-Based Quantum Computers}
\author{Ari Mizel}
\affiliation{Physics Department and Materials Research Institute, Pennsylvania State University, University Park, PA 16802}
\author{Daniel A. Lidar}
\affiliation{Chemical Physics Theory Group, University of Toronto, 80 St. George St.,
Toronto, Ontario M5S 3H6, Canada}

\begin{abstract}
In the effort to design and to construct a quantum computer, several
leading proposals make use of spin-based qubits. These designs
generally assume that spins undergo pairwise interactions. We point out
that, when several spins are engaged mutually in pairwise
interactions, the quantitative strengths of the interactions can
change and qualitatively new terms can arise in the Hamiltonian,
including four-body interactions. In parameter regimes of experimental
interest, these coherent effects are large enough to interfere with
computation, and may require new error correction or avoidance techniques.

\end{abstract}

\maketitle

Researchers in the field of quantum computation seek to exploit
quantum coherence to speed calculation.  While theoretical
possibilities have been enticing, the prospect of realizing a
practical quantum computer is quite daunting.  A number of physical
systems have been suggested as candidate instantiations.  One
particularly promising proposed design \cite{Loss:98} involves
electron spins localized in quantum dots.  In its original form, the
proposal supplements tunable exchange interactions between electron
spins with single-qubit operations to achieve a universal quantum
computer.  
Subsequent research has shown that a universal set of gates
can be constructed using the exchange interaction alone, provided that
one encodes a logical qubit into the state of several spins
(``encoded universality'') \cite{Bacon:99a}.
Motivated by the scheme of Ref.~\cite{Loss:98}, there have been a
number of studies of the one-particle and two-particle behavior of
electrons localized on quantum dots within a quantum computer
\cite{Burkard:99,Burkard:00}. Starting from the simplest case of two
electrons in singly-occupied dots in the lowest orbital state,
systematic generalizations have been introduced and their effect on
the exchange interaction studied.  In particular, researchers have
analyzed the effect of double occupation
\cite{Hu:99,Schliemann:01,Barrett:02}, higher orbital states
\cite{Burkard:99,Hu:99}, many-electron dots \cite{Hu:01b}, and
spin-orbit coupling \cite{Kavokin:01}.
Here we point out that many-body interactions
can appear once there are more than two
electron spins in a computer.  In a computer with at least three
coupled dots containing three electron spins, multi-particle exchange
processes can change the strength of exchange interactions.  In a
computer with four coupled dots containing four electron spins,
four-body interactions can arise.
Such modifications to the exchange
interactions have not been previously addressed in the quantum
computing literature, and
require careful consideration, and possibly removal by error correction
\cite{Gottesman:97a} or
avoidance \cite{Bacon:99a},
if one intends to scale a
quantum computer beyond two spins.  Their ramifications for quantum
computation may also extend to (i) the encoded universality paradigm,
where in efficient implementations several exchange interactions
are turned on simultaneously \cite{Bacon:99a};
(ii) adiabatic quantum computing \cite{Farhi:01}, where the final 
 Hamiltonian for any non-trivial calculation inevitably includes 
 simultaneous interactions between multiple qubits;
(iii) fault-tolerant quantum error correction, where a higher degree of
parallelism translates into a lower threshold for fault-tolerant
quantum computation operations
\cite{Gottesman:97a}; (iv)
the ``one-way'' quantum computer proposal,
where all nearest-neighbor interactions in a cluster of coupled spins are
turned on simultaneously in order to prepare a many-spins entangled state \cite{Raussendorf:01}; (v)
the search for physical systems with intrinsic, topological fault
tolerance, where systems with four-body interactions have recently
been identified as having the sought-after properties \cite{Freedman:02}.

Our analysis begins with the derivation of an effective Hamiltonian
for the electron spins in the quantum computer.
The microscopic Hamiltonian describing Coulomb-coupled electrons is
\begin{equation}
H = \sum_{i=1}^{n}\frac{\mathbf{p}_{i}^{2}}{2m}+V(\mathbf{r}_{i})+\sum_{i<j} \frac{e^{2}}{\kappa |\mathbf{r}_{i}-\mathbf{r}_{j}|},
\label{spatialHamiltonian}
\end{equation}
where the confining potential $V(\mathbf{r})$ contains $n$ energy
minima, which give rise to the $n$ quantum dots that house the
electrons.  We assume at first that each dot contains a single energetically
accessible orbital, in accordance with the Heitler-London (HL)
approximation \cite{Heitler:27}, and we label the orbitals with
capital letters $A,B,\ldots$  Since electrons have two possible spin
states, this assumption leaves the $n$-electron system with $2^n$
basis states
\begin{equation*}
\left\vert \Psi (s_{A},s_{B},\dots)\right\rangle = \sum_{P}\delta
_{P}P[\left\vert AB\dots \right\rangle \left\vert s_{A}s_{B} \dots
\right\rangle ].
\end{equation*}
Here, the first ket $\left\vert A B\dots \right\rangle$ refers to the
orbital states of the electrons and the second ket $\left\vert
s_{A}s_{B} \dots \right\rangle$ indicates the spin of each electron.
The sum runs over all $n!$ permutations $P$, where $\delta _{P}=1$
$(-1)$ if the permutation is even (odd), ensuring overall
antisymmetry.  In every term of the sum, the electron in orbital $A$ has spin $s_A$.

In this $2^{n}$ dimensional basis, the Hamiltonian
(\ref{spatialHamiltonian}) takes the form of a $2^{n}\times 2^{n}$
Hermitian matrix, specified by $4^{n}$ real numbers. One obtains an
electron-spin representation of the Hamiltonian matrix by writing it
as a sum of $4^{n}$ Hermitian spin matrices of the form
$\sigma_{i,j,\cdots} \equiv \sigma _{i}(A)\otimes \sigma
_{j}(B)\otimes \cdots$ each multiplied by a real coefficient (there
are $n$ factors in $\sigma_{i,j,\cdots}$). Here, $\sigma _{i}(p)$
denotes the Pauli matrix $\sigma _{i}$ acting on the electron in dot
$p $, with $i =0,1,2,3$ and with $\sigma _{0}$ equal to the identity
matrix.  This decomposition into spin matrices produces an effective
electron-spin Hamiltonian that conveniently describes the dynamics of
$n$ qubits.

Symmetry considerations can fundamentally constrain the form of the
electron-spin Hamiltonian. For simplicity, we will assume that the
quantum dots in the computer are arranged in a completely symmetric
fashion (i.e., an equilateral triangle for 3 electrons, an equilateral
tetrahedron for 4 electrons).  We will also neglect spin-orbit
coupling and any external magnetic field.  These assumptions are
introduced for convenience -- they are not essential to our
conclusions, and will be relaxed in a future publication \cite{tbp}.
They simplify the analysis by implying that the
effective spin operator Hamiltonian has rotation, inversion, and
exchange symmetry. It can therefore only be a function of the
magnitude
squared $\mathbf{S}_{T}^{2}=(\mathbf{S}
_{A}+\mathbf{S}_{B}+\dots )^{2}$ of the total spin
$\mathbf{S}_{T}$. We must have
\begin{equation}
H_{\mathrm{spin}}=L_{0}+L_{1}\mathbf{S}_{T}^{2}+L_{2}(\mathbf{S}
_{T}^{2})^{2}+\dots
\label{spinHamiltonian}
\end{equation}
where $L_{0},L_{1},L_{2},\dots $ are real constants with dimensions of
energy (we take spin to be dimensionless). The
constant $L_{0}$ is an energy shift. The term proportional to $ L_{1}$
gives rise to the familiar Heisenberg interaction. Here we see that in
principle higher order interactions may be present in the spin
Hamiltonian, starting with a fourth order term proportional to
$L_{2}$.  (The constants $L_{n}$ are expected to decrease in magnitude
with $n$ but we will demonstrate that at least $L_2$ can be physically
important in quantum computations.)

To compute the values of $L_0 ,L_1 ,L_2 ,\dots $ of the $n$ electron
effective-spin Hamiltonian, we consider an eigenstate $ |\Psi
_{S_{T}}^{n} \rangle $ of $\mathbf{S}_{T}^{2}$, with known eigenvalue
$ S_{T}(S_{T}+1)$. We compute the expectation value of the effective
spin Hamiltonian (\ref{spinHamiltonian}) in this state,
compute the expectation value of the spatial Hamiltonian
(\ref{spatialHamiltonian}), and equate the two:
\begin{equation}
\langle \Psi _{S_{T}}^{n} |H_{\mathrm{spin}}|\Psi _{S_{T}}^{n} \rangle =\langle \Psi _{S_{T}}^{n}|H|\Psi _{S_{T}}^{n} \rangle .
\label{eq:equal}
\end{equation}
This procedure is repeated for all eigenvalues of $\mathbf{S}_{T}^{2}$, thus
generating a set of linear equations for the parameters $L_0 ,L_1 ,L_2
,\dots $, in terms of matrix elements of $H$ between different orbital
states. For $n$ electrons the number of distinct eigenvalues of $\mathbf{S}
_{T}^{2}$ is $\lfloor \frac{n}{2} \rfloor+1 $ (where $\lfloor \frac{n}{2}
\rfloor$ denotes the greatest integer less than $\frac{n}{2})$, so this is
the maximum number of distinct energy eigenvalues of the Hamiltonian (\ref{spinHamiltonian}). The Hamiltonian need only contain this many degrees of
freedom to fix all of its eigenvalues, so we can set $L_m = 0$ for
$m \ge \lfloor \frac{n}{2} \rfloor + 1$. We are led to $\lfloor \frac{n}{2}
\rfloor+1$ coupled linear equations for the non-zero $L_m$ parameters.

In the three electron case, the total spin can be $S_{T}=1/2$ or
$S_{T}=3/2$. We therefore need to solve $\lfloor \frac{3}{2}\rfloor
+1=2$ equations, and it is sufficient to keep only two constants $L_0
$ and $L_1$ in $H_{\mathrm{spin}}$, setting
$L_{m\ge 2}=0$. As a convenient state with known $S_{T}=3/2$ we take the
normalized state
$|\Psi _{3/2}^{3}\rangle \propto \left\vert \Psi
(\uparrow \uparrow \uparrow )\right\rangle $.  Equation~(\ref{eq:equal}) gives
\begin{equation}
E_{3/2} = L_0 +\frac{15}{4}L_1 =\frac{\epsilon_{3}+2\epsilon_{0}-3
  \epsilon_{1}}{p_{3}+2p_{0}-3p_{1}}.
\label{ST3/2}
\end{equation}
Here, we have defined 
\begin{eqnarray*}
p_{3}=\left\langle ABC\right\vert ABC\rangle, && \epsilon_{3} =\left\langle ABC\right\vert H\left\vert ABC\right\rangle , \\
p_{1}=\left\langle BAC\right\vert ABC\rangle, && \epsilon_{1} =\left\langle BAC\right\vert H\left\vert ABC\right\rangle, \\
p_{0}=\left\langle CAB\right\vert ABC\rangle, && \epsilon_{0}
=\left\langle CAB\right\vert H\left\vert ABC\right\rangle ,
\end{eqnarray*}
where the subscript $i$ in $p_{i}$ or $\epsilon_{i}$ denotes the
number of electrons with the same orbital state in bra and ket.  For
the case $S_{T}=1/2$, using
$|\Psi _{1/2}^{3}\rangle \propto
\frac{1}{\sqrt{2}} \left( \left\vert \Psi (\uparrow \downarrow
\uparrow )\right\rangle -\left\vert \Psi (\downarrow \uparrow \uparrow
)\right\rangle \right)$ yields:
\begin{equation}
E_{1/2} = L_0 +\frac{3}{4}L_1 =\frac{\epsilon_{3}-\epsilon_{0}}{p_{3}-p_{0}}.
\label{ST1/2}
\end{equation}
To compute the usual exchange coupling, it is useful to rewrite
$H_{\mathrm{ spin}}$ as
\begin{eqnarray}
H_{\mathrm{spin}} &=&(L_0 +L_1 \sum_{A\leq i\leq C} \mathbf{S} _{i}^{2})+2L_1
\sum_{A\leq i<j \leq C} \mathbf{S}_{i}\cdot \mathbf{S}_{j}  \notag \\
&\equiv& K + J(\mathbf{S}_{A}\cdot \mathbf{S}_{B}+\mathbf{S} _{A}\cdot 
\mathbf{S}_{C}+\mathbf{S}_{B}\cdot \mathbf{S}_{C}). \label{eq:Hspin3}
\end{eqnarray}
The constants $K = L_0 +\frac{9}{4}L_1$ and $J = 2L_1$ can be expressed
in terms of the $p_{i}$ and $\epsilon_{i}$
using Eqs.~(\ref {ST3/2})
and (\ref{ST1/2}).  Note that the value of the exchange constant $J$
is determined in part by the ``three-electron-exchange'' terms of the form
$p_{0}=\left\langle CAB\right. \left\vert ABC\right\rangle$ and
$\epsilon_{0}=\left\langle CAB\right\vert H\left\vert ABC\right\rangle$.  Such terms involve a cooperative effect between all three
electrons and hence cannot be seen in two-electron calculations.
Thus, \emph{the presence of the third electron quantitatively changes
the exchange coupling between the other two electrons}.

To compute values for the $p_i$ and $\epsilon _i$, it is necessary to select a
specific form for the one-body potential in (\ref{spatialHamiltonian}).  We choose the sample potential 
\begin{equation}  
V(\mathbf{r}) = \frac{1}{2(2l)^6} m \omega _o ^2 |\mathbf{r} - \mathbf{A}|^2
|\mathbf{r} - \mathbf{B}|^2 |\mathbf{r} - \mathbf{C}|^2 |\mathbf{r} - 
\mathbf{D}|^2.
\label{V}
\end{equation}
This potential has a quadratic minimum at each of the vertices of an
equilateral tetrahedron $\mathbf{A} = (0,0,0)$, $\mathbf{B} =
(2l\sqrt{\frac{ 1}{3}},0,-2l\sqrt{\frac{2}{3}})$, $\mathbf{C} =
(-l\sqrt{\frac{1}{3}},l,-2l \sqrt{\frac{2}{3}})$, and $\mathbf{D} =
(-l\sqrt{\frac{1}{3}},-l,-2l\sqrt{ \frac{2}{3}})$, which become the
locations of the dots $A,B,C,D$. The distance between vertices is
$2l$. We select a potential with four minima so that it can be used in
the four electron case without modification; the extra
minimum does not influence the three- electron case in any significant
way.  At vertex $\mathbf{A}$, we define the localized Gaussian state
$\phi _A (\mathbf{r}) \equiv \langle \mathbf{r}|A\rangle \equiv \left(\frac{m
\omega _o}{\pi \hbar} \right)^{3/4} \exp \left( -\frac{m \omega _o}{2 \hbar}
|\mathbf{r}-\mathbf{A}|^2 \right)$
which is the ground state of the quadratic minimum at that vertex. We
define localized states similarly for the other vertices.  The
Gaussian form makes it possible to obtain all $p_i$ and $\epsilon_i$
analytically in terms of the energy $\hbar \omega _o$, and the
dimensionless parameters $x_b \equiv m\omega _{o}l^{2}/\hbar$ and
$x_c \equiv e^{2}/(\kappa l\hbar \omega _{o})$.
Physically, $x_b$ is the ratio of the tunneling energy barrier $\frac{1}{2}m
\omega_o^2 l^2$ to the harmonic oscillator ground state energy $E_g=\frac{1}{2}
\hbar  \omega_o$, while $x_c$ is the Coulomb energy over $E_g$.

\begin{figure}[tbp]
  \includegraphics[height=5.3cm,angle=0]{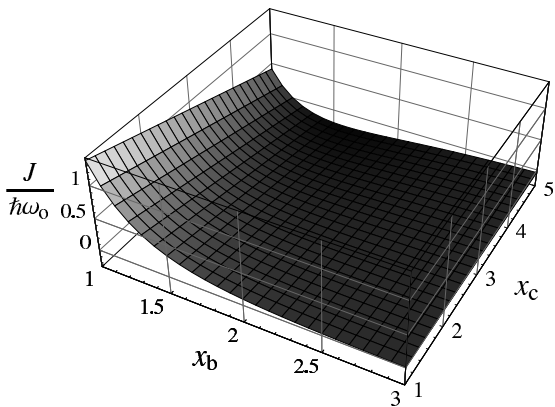}
\caption{Plot of the exchange constant as a function of tunneling barrier and
  Coulomb energy in the case of three mutually interacting electrons
  (dimensionless units).}
\label{J3}
\end{figure}

In Fig.~\ref{J3}, we plot the exchange-interaction constant $J$ in
units of $\hbar \omega_o$ as a function of $x_b$ and $b_c$.  The plot
generally indicates that $J$ increases as the tunneling barrier
decreases ($x_b$ smaller), an intuitively reasonable result.  A negative
minimum is visible when $x_c \sim 5$; this may reflect the breakdown of
the HL approximation in the regime of strong interactions.  Following
Ref.~\onlinecite{Loss:98}, we estimate that realistic values for GaAs
heterostructure single dots, $x_b \sim 1$, $x_c \sim 1.5$, and $\hbar
\omega_o = 3\; \mathrm{meV}$, do not fall in this strong interactions
regime.

\begin{figure}[tbp]
\includegraphics[height=5.3cm,angle=0]{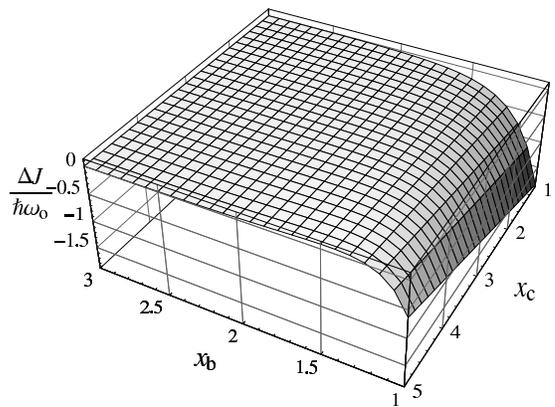} 
\caption{Plot of the change in the exchange constant due to
  $3$-electron-swap matrix elements. Horizontal axes as in Fig.~\ref{J3} (with their directions flipped).}
\label{dJ3}
\end{figure}

Fig.~\ref{dJ3} shows the change $\Delta J$ that results from
three-electron-swap matrix elements $ \epsilon _0$ and
$p_0$ (i.e. $\Delta J$ is the change in $J = 2L_1$ when $\epsilon _0$
and $p_0$ are set to zero in expressions (\ref{ST3/2}) and
(\ref{ST1/2})).  Comparing the scales of Figs.~\ref{J3} and \ref{dJ3},
one finds that the three-electron swap matrix elements can have a
powerful influence on $J$.  They will strongly impact quantum
evolution when two-particle gates act simultaneously on three qubits
(as may arise in circumstances (i) - (iii) above).

In the four electron case, a still more striking effect arises.  Here,
it is possible to have $S_{T}=0$, $S_{T}=1$, or $S_{T}=2$, so that one
keeps three constants $L_0 $, $L_1 $, and $L_2$ in
$H_{\mathrm{spin}}$. It follows immediately that $H_{\mathrm{spin}} $
includes terms of the form $L_2 (\mathbf{S}_{A}\cdot \mathbf{S}_{B})(
\mathbf{S}_{C}\cdot \mathbf{S}_{D})$ and permutations. Unless $L_2$
happens to vanish, \emph{the presence of a fourth electron introduces
a qualitatively new 4-body interaction as well as a quantitative
change in the exchange coupling between the other electrons.}

We evaluate $L_0$, $L_1$, and $L_2$ using Eq.~(\ref{eq:equal}).
Extending the three-electron definitions, we set
\begin{eqnarray*}
p_{0} =\left\langle BADC\right\vert ABCD\rangle, &&
\epsilon_{0}=\left\langle BADC\right\vert H|ABCD\rangle, \\
p_{0}^{\prime } =\left\langle DABC\right\vert ABCD\rangle, &&
\epsilon_{0}^{\prime }=\left\langle DABC\right\vert H\left\vert
ABCD\right\rangle, \\
p_{1} =\left\langle ADBC\right\vert ABCD\rangle, &&
\epsilon_{1}=\left\langle ADBC\right\vert H|ABCD\rangle, \\
p_{2} =\left\langle BACD\right\vert ABCD\rangle, &&
\epsilon_{2}=\left\langle BACD\right\vert H|ABCD\rangle, \\
p_{4} =\left\langle ABCD\right\vert ABCD\rangle, &&
\epsilon_{4}=\left\langle ABCD|H\right\vert ABCD\rangle.
\end{eqnarray*}

A convenient state to use for
$S_{T}=0$ is $|\Psi^4_0\rangle \propto \left( \left\vert \Psi
(\uparrow \downarrow \uparrow \downarrow )\right\rangle -\left\vert
\Psi (\uparrow \downarrow \downarrow \uparrow )\right\rangle
-\left\vert \Psi (\downarrow \uparrow \uparrow \downarrow
)\right\rangle +|\Psi (\downarrow \uparrow \downarrow \uparrow
)\rangle \right) $. After normalization, this state yields the singlet
energy
\begin{equation*}
E_{0}=L_0 =\frac{\epsilon_{4}-4\epsilon_{1}+3\epsilon_{0}}{
  p_{4}-4p_{1}+3p_{0}}.
\label{ST0}
\end{equation*}
A convenient state to use for $S_{T}=1$ is $|\Psi^4_1\rangle \propto
\left( \left\vert \Psi (\uparrow \downarrow \uparrow \downarrow
)\right\rangle +\left\vert \Psi (\uparrow \downarrow \downarrow \uparrow
)\right\rangle -\left\vert \Psi (\downarrow \uparrow \uparrow \downarrow
)\right\rangle -|\Psi (\downarrow \uparrow \downarrow \uparrow )\rangle
\right) $. This state, after normalization, yields the triplet energy 
\begin{equation*}
E_{1}=L_0 +2L_1 +4L_2 =\frac{\epsilon_{4}-2\epsilon_{2}-\epsilon_{0}+2
  \epsilon_{0}^{\prime }}{p_{4}-2p_{2}-p_{0}+2p_{0}^{\prime }}
\label{ST1}
\end{equation*}
Finally, a convenient state to use for $S_{T}=2$ is $|\Psi^4_2\rangle \propto \left\vert \Psi
(\uparrow \uparrow \uparrow \uparrow )\right\rangle .$ We find for the
quintet energy 
\begin{equation*}
E_{2}=L_0 +6L_1 +36L_2 =\frac{\epsilon_{4}-6\epsilon_{2}+8\epsilon_{1}+3
\epsilon_{0}-6\epsilon_{0}^{\prime }}{p_{4}-6p_{2}+8p_{1}+3p_{0}-6p_{0}^{\prime }}.
\label{ST2}
\end{equation*}

We would like to exhibit interaction constants explicitly in the spin
Hamiltonian. Spin-operator identities 
allow one to convert from the form (\ref{spinHamiltonian}) to
\begin{eqnarray*}
H_{\mathrm{spin}} &=& K+J\sum_{i<j}\mathbf{S}_{i}\cdot \mathbf{S}
_{j}+J^{\prime } [ \left( \mathbf{S}_{A}\cdot \mathbf{S}_{B}\right)
\left( \mathbf{S}_{C}\cdot \mathbf{S}_{D}\right) \\ &+& \left(
\mathbf{S} _{A}\cdot \mathbf{S}_{C}\right) \left( \mathbf{S}_{B}\cdot
\mathbf{S} _{D}\right) + \left( \mathbf{S}_{A}\cdot
\mathbf{S}_{D}\right) \left( \mathbf{S}_{B}\cdot \mathbf{S}_{C}\right)
]
\end{eqnarray*}
where $K = L_0 +3L_1 +\frac{27}{2}L_2$, $J= 2L_1 + 14L_2$, and
$J^{\prime } = 8L_2$.  Generically, $J^{\prime }$ does not vanish, and
four-body interactions arise.

\begin{figure}[tbp]
\includegraphics[height=5.2cm,angle=0]{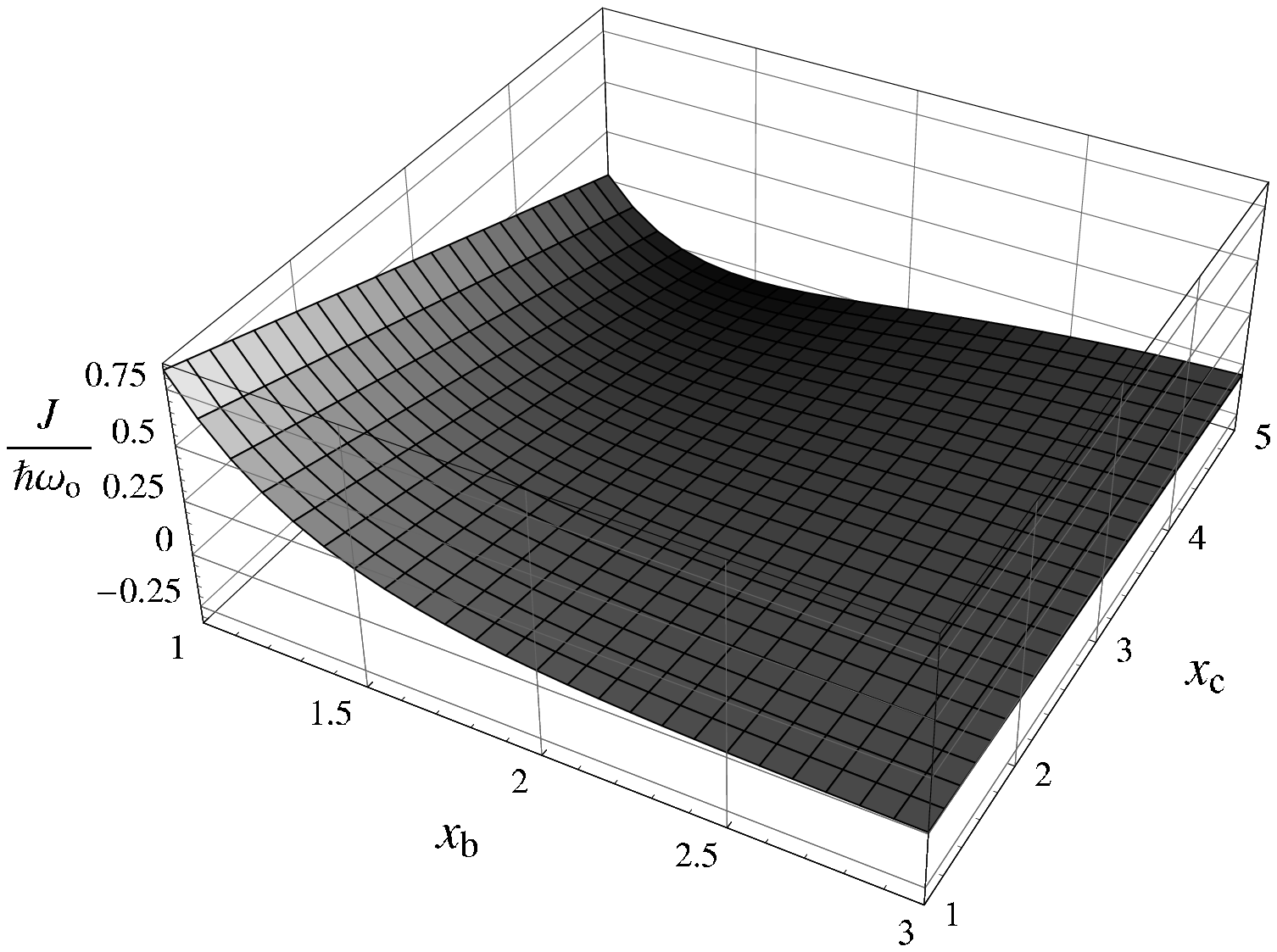} 
\caption{Plot of the exchange constant in the case of four mutually interacting electrons. Horizontal axes as in Fig.~\ref{J3}.}
\label{J4}
\end{figure}

\begin{figure}[tbp]
\includegraphics[height=5.2cm,angle=0]{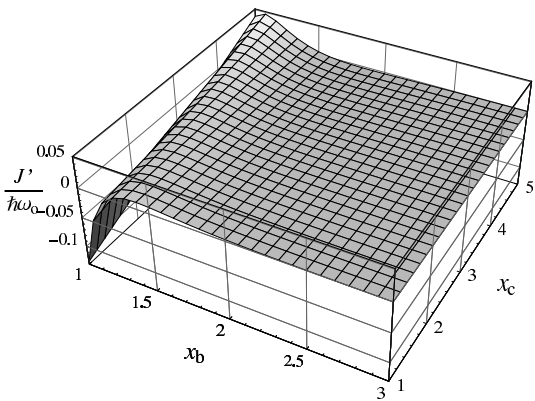} 
\caption{Plot of the $4$-body exchange constant in the case of four
  mutually interacting electrons. Horizontal axes as in Fig.~\ref{J3}.}
\label{Jp4}
\end{figure}

We use the potential (\ref{V}) to estimate the magnitude of the
effect.  The exchange-interaction constant $J$ is plotted in
Fig.~\ref{J4}.  Its form is similar to that of $J$ in the
three-electron case, with a slightly reduced magnitude.  The four-body
interaction constant, $J^{\prime}$, appears in Fig.~\ref{Jp4}.  In the
parameter region of likely experimental interest ($x_b \sim 1$, $x_c \sim
1.5$, $\hbar \omega_o \sim 3\; \mathrm{meV}$), $J^\prime / J \sim -15
\%$, certainly large enough to demand attention in computer design.
Experimentally, four-body terms have been observed in $^3$He
\cite{Roger:83}, and Cu$_4$O$_4$ square plaquettes in La$_2$CuO$_4$
\cite{Coldea:01}, where $J^{\prime}/J$ was found to be $\sim 27\% $.

Finally, to test the accuracy of our HL calculations, we have used a Hund-Mulliken (HM) calculation that extends the
HL basis to include basis states with double occupation: two electrons
on a single dot. For instance, the HM Hilbert space describing three
electrons on three dots splits into a four-dimensional total-spin
subspace with $S_{T}=3/2$ (same as in the HL case), and a
$16$-dimensional $S_{T}=1/2$ subspace, comprised of two degenerate
$S_{T,z}=\pm 1/2$ subspaces. For parameters of interest
\cite{tbp}, twelve $S_{T}=1/2$ states have high energy and involve
substantial double occupation while the remaining four $S_{T} = 1/2$
states form a degenerate ground state similar to that of the HF basis.  Therefore one recovers the HF picture by projecting out
the $8$ lowest members of the HM solution (the four low $S_{T} = 1/2$
and the four $S_{T} = 3/2$ states) to obtain an exchange
Hamiltonian (\ref{eq:Hspin3}).  The corresponding $J$ qualitatively confirms the
form of Fig.~\ref{J3} and Fig.~\ref{dJ3}.  Thus, even when
HL results are not necessarily quantitatively precise, substantial
many-body corrections to the interaction Hamiltonian persist in more
accurate calculations.  In the context of quantum computation, these
effects could, on one hand, utterly derail an algorithm or, on the
other, find uses in novel designs.

\begin{acknowledgments}
A.M. acknowledges the support of the Packard Foundation.
D.A.L. acknowledges support under the DARPA-QuIST program (managed by
AFOSR under agreement No. F49620-01-1-0468), and the Connaught
Fund. We thank Prof. T.A. Kaplan for useful correspondence.
\end{acknowledgments}


\begin{thebibliography}{43}
\expandafter\ifx\csname natexlab\endcsname\relax\def\natexlab#1{#1}\fi
\expandafter\ifx\csname bibnamefont\endcsname\relax
  \def\bibnamefont#1{#1}\fi
\expandafter\ifx\csname bibfnamefont\endcsname\relax
  \def\bibfnamefont#1{#1}\fi
\expandafter\ifx\csname citenamefont\endcsname\relax
  \def\citenamefont#1{#1}\fi
\expandafter\ifx\csname url\endcsname\relax
  \def\url#1{\texttt{#1}}\fi
\expandafter\ifx\csname urlprefix\endcsname\relax\def\urlprefix{URL }\fi
\providecommand{\bibinfo}[2]{#2}
\providecommand{\eprint}[2][]{\url{#2}}

\bibitem[{\citenamefont{{D. Loss and D.P. DiVincenzo}}(1998)}]{Loss:98}
\bibinfo{author}{\bibnamefont{{D. Loss and D.P. DiVincenzo}}},
  \bibinfo{journal}{Phys. Rev. A} \textbf{\bibinfo{volume}{57}},
  \bibinfo{pages}{120} (\bibinfo{year}{1998}).

\bibitem[{\citenamefont{{D. Bacon, J. Kempe, D.A. Lidar and K.B.
  Whaley}}(2000)}]{Bacon:99a}
\bibinfo{author}{\bibnamefont{{D. Bacon, J. Kempe, D.A. Lidar and K.B.
  Whaley}}}, \bibinfo{journal}{Phys. Rev. Lett.} \textbf{\bibinfo{volume}{85}},
  \bibinfo{pages}{1758} (\bibinfo{year}{2000});
\bibinfo{author}{\bibnamefont{{J. Kempe, D. Bacon, D.A. Lidar, and K.B.
  Whaley}}}, \bibinfo{journal}{Phys. Rev. A} \textbf{\bibinfo{volume}{63}},
  \bibinfo{pages}{042307} (\bibinfo{year}{2001});
\bibinfo{author}{\bibnamefont{{D.P. DiVincenzo, D. Bacon, J. Kempe, G. Burkard,
  and K.B. Whaley}}}, \bibinfo{journal}{Nature} \textbf{\bibinfo{volume}{408}},
  \bibinfo{pages}{339} (\bibinfo{year}{2000});
\bibinfo{author}{\bibnamefont{{D.A. Lidar and L.-A. Wu}}},
  \bibinfo{journal}{Phys. Rev. Lett.} \textbf{\bibinfo{volume}{88}},
  \bibinfo{pages}{017905} (\bibinfo{year}{2002}).

\bibitem[{\citenamefont{{G. Burkard, D. Loss and D.P.
  DiVincenzo}}(1999)}]{Burkard:99}
\bibinfo{author}{\bibnamefont{{G. Burkard, D. Loss and D.P. DiVincenzo}}},
  \bibinfo{journal}{Phys. Rev. B} \textbf{\bibinfo{volume}{59}},
  \bibinfo{pages}{2070} (\bibinfo{year}{1999}).

\bibitem[{\citenamefont{{G. Burkard, H.-A. Engel and D.
  Loss}}(2000)}]{Burkard:00}
\bibinfo{author}{\bibnamefont{{G. Burkard, H.-A. Engel and D. Loss}}},
  \bibinfo{journal}{Fortschr. Phys.} \textbf{\bibinfo{volume}{48}},
  \bibinfo{pages}{965} (\bibinfo{year}{2000}).

\bibitem[{\citenamefont{{X. Hu and S. Das Sarma}}(2000)}]{Hu:99}
\bibinfo{author}{\bibnamefont{{X. Hu and S. Das Sarma}}},
  \bibinfo{journal}{Phys. Rev. A} \textbf{\bibinfo{volume}{61}},
  \bibinfo{pages}{062301} (\bibinfo{year}{2000}).

\bibitem[{\citenamefont{{J. Schliemann, D. Loss, and A.H.
  MacDonald}}(2001)}]{Schliemann:01}
\bibinfo{author}{\bibnamefont{{J. Schliemann, D. Loss, and A.H. MacDonald}}},
  \bibinfo{journal}{Phys. Rev. B} \textbf{\bibinfo{volume}{63}},
  \bibinfo{pages}{085311} (\bibinfo{year}{2001}).

\bibitem[{\citenamefont{{S.D. Barrett and C.H.W. Barnes}}(2002)}]{Barrett:02}
\bibinfo{author}{\bibnamefont{{S.D. Barrett and C.H.W. Barnes}}},
  \bibinfo{journal}{Phys. Rev. B} \textbf{\bibinfo{volume}{66}},
  \bibinfo{pages}{123518} (\bibinfo{year}{2002}).

\bibitem[{\citenamefont{{X. Hu and S. Das Sarma}}(2001)}]{Hu:01b}
\bibinfo{author}{\bibnamefont{{X. Hu and S. Das Sarma}}},
  \bibinfo{journal}{Phys. Rev. A} \textbf{\bibinfo{volume}{64}},
  \bibinfo{pages}{042312} (\bibinfo{year}{2001}).

\bibitem[{\citenamefont{{K.V. Kavokin}}(2001)}]{Kavokin:01}
\bibinfo{author}{\bibnamefont{{K.V. Kavokin}}}, \bibinfo{journal}{Phys. Rev. B}
  \textbf{\bibinfo{volume}{64}}, \bibinfo{pages}{075305}
  (\bibinfo{year}{2001}).

\bibitem[{\citenamefont{Gottesman}(1997)}]{Gottesman:97a}
\bibinfo{author}{\bibfnamefont{D.}~\bibnamefont{Gottesman}},
  \bibinfo{journal}{Phys. Rev. A} \textbf{\bibinfo{volume}{57}},
  \bibinfo{pages}{127} (\bibinfo{year}{1997});
\bibinfo{author}{\bibfnamefont{J.}~\bibnamefont{Preskill}},
  \bibinfo{journal}{Proc. Roy. Soc. London Ser. A}
  \textbf{\bibinfo{volume}{454}}, \bibinfo{pages}{385} (\bibinfo{year}{1998});
\bibinfo{author}{\bibnamefont{{A.M. Steane}}}, \bibinfo{journal}{Nature}
  \textbf{\bibinfo{volume}{399}}, \bibinfo{pages}{124} (\bibinfo{year}{1999});
\bibinfo{author}{\bibnamefont{{D.A. Lidar, D. Bacon, J. Kempe, and K.B.
Whaley}}}, \bibinfo{journal}{Phys. Rev. A} \textbf{\bibinfo{volume}{63}},
\bibinfo{pages}{022307} (\bibinfo{year}{2001}).

 \bibitem[{\citenamefont{{E. Farhi, J. Goldstone, S. Gutmann, J. Lapan, 
 A. Lundgren, D. Preda}}(2001)}]{Farhi:01}
 \bibinfo{author}{\bibnamefont{{E. Farhi, J. Goldstone, S. Gutmann, J. 
 Lapan, A. Lundgren, D. Preda}}},
  \bibinfo{journal}{Science} \textbf{\bibinfo{volume}{292}},
  \bibinfo{pages}{472} (\bibinfo{year}{2001}).

\bibitem[{\citenamefont{{R. Raussendorf and H.J.
  Briegel}}(2001)}]{Raussendorf:01}
\bibinfo{author}{\bibnamefont{{R. Raussendorf and H.J. Briegel}}},
  \bibinfo{journal}{Phys. Rev. Lett.} \textbf{\bibinfo{volume}{86}},
  \bibinfo{pages}{5188} (\bibinfo{year}{2001}).

\bibitem[{\citenamefont{{M.H. Freedman}}(2002)}]{Freedman:02}
\bibinfo{author}{\bibnamefont{{M.H. Freedman}}}, \eprint{quant-ph/0110060}.

\bibitem[{\citenamefont{{W. Heitler and F. London}}(1927)}]{Heitler:27}
\bibinfo{author}{\bibnamefont{{W. Heitler and F. London}}},
  \bibinfo{journal}{Z. Physik} \textbf{\bibinfo{volume}{44}},  \bibinfo{pages}{455} (\bibinfo{year}{1927}).

\bibitem[{\citenamefont{{A.Mizel and D.A. Lidar}}(2003)}]{tbp}
\bibinfo{author}{\bibnamefont{{A. Mizel and D.A. Lidar}}},
\bibinfo{note}{in preparation}.

\bibitem[{\citenamefont{{M. Roger}}(1983)}]{Roger:83}
\bibinfo{author}{\bibnamefont{{M. Roger, J.H. Hetherington, and J.M. Delrieu}}},
  \bibinfo{journal}{Rev. Mod. Phys.} \textbf{\bibinfo{volume}{55}},
  \bibinfo{pages}{1} (\bibinfo{year}{1983}).

\bibitem[{\citenamefont{{R. Coldea, S.M. Hayden, G. Aeppli,
      T.G. Perring, C.D. Frost, T.E. Mason, S.-W. Cheong, and Z. Fisk}}(2001)}]{Coldea:01}
\bibinfo{author}{\bibnamefont{{R. Coldea, S.M. Hayden, G. Aeppli,
      T.G. Perring, C.D. Frost, T.E. Mason, S.-W. Cheong, and Z. Fisk}}},
  \bibinfo{journal}{Phys. Rev. Lett.} \textbf{\bibinfo{volume}{86}},
  \bibinfo{pages}{5377} (\bibinfo{year}{2002}).

\end{thebibliography}
\end{document}